# An integrated *in vitro* – *in silico* approach for silver nanoparticle dosimetry in cell cultures

Daniele Poli[1], Giorgio Mattei[2], Nadia Ucciferri[1] and Arti Ahluwalia[1, 2*]

[1]Research Center E. Piaggio, University of Pisa, Pisa, Italy; [2]Department of Information Engineering, University of Pisa, Pisa, Italy

*Corresponding author: Arti Ahluwalia (arti.ahluwalia@unipi.it)

## Abstract

Potential human and environmental hazards resulting from the exposure of living organisms to silver nanoparticles (Ag NPs) have been the subject of intensive discussion in the last decade. Despite the growing use of Ag NPs in biomedical applications, a quantification of the toxic effects as a function of the total silver mass reaching cells (namely, target cell dose) is still needed. To provide a more accurate dose-response analysis, we propose a novel integrated approach combining well-established computational and experimental methodologies. We first used the particokinetic model (ISD3) proposed by Thomas and colleagues (2018) for providing experimental validation of computed Ag NP sedimentation in static-cuvette experiments. After validation, ISD3 was employed to predict the total mass of silver reaching human endothelial cells and hepatocytes cultured in 96 well plates. Cell viability measured after 24h of culture was then related to this target cell dose. Our results show that the dose perceived by the cell monolayer after 24 h of exposure is around 85% lower than the administered nominal media concentration. Therefore, accurate dosimetry considering particle characteristics and experimental conditions (e.g., time, size and shape of wells) should be employed for better interpreting effects induced by the amount of silver reaching cells.

**Introduction**

Engineered nanomaterials (ENMs) are very successful in the bio-technology industry because of their exceptionally small size and unique physical and chemical properties [12,13]. One of the most widely used ENMs is silver, popularly known for its antimicrobial properties. It can be coated on biomedical devices [9], used in medical contexts for personal health care [16] or biological applications [35], and adapted for food products such as kitchen tools, storage containers and cutting boards [30]. However, despite the growing use of silver nanoparticles (Ag NPs) in the last decade [1,7,25,43], potential human and environmental hazards resulting from exposure to Ag NPs continue to be the subject of attention [37,52] owing to their well-documented toxicity both *in vivo* and *in vitro*.

Le has succinctly discussed and summarised the increased exposure to Ag NPs and the possible effects related to their short- and long-term toxicity (e.g. decreased cell viability and apoptosis) [25]. Well-established *in vivo* models [6,11,45], as well as *in vitro* systems [32,44,48], have been also proposed for toxicological studies. Mouse and zebrafish models exposed to ENMs, for example, showed nanotoxicity effects on female reproductive and fetal development [26,39]. *In vitro*, dose-dependent Ag NPs induced cellular necrosis, inflammation and oxidative stress in living organisms in a size-specific manner [5,15,23,29,33,49]. The oxidative stress was further related to the anti-microbial activity of Ag NPs affecting different types of pathogens [8,14,38]. Genetic damage (e.g., DNA breakage) was additionally found within the cells interacting with Ag NPs [18] and associated with the production of reactive oxygen species [34,42].

However, toxicity results are often misinterpreted since they are generally reported as a function of the initial silver nanoparticle concentration present in the culture medium (here referred to as *nominal media concentration*) and not as a function of that actually coming into contact with cells (i.e. *target cell dose*, here defined as the total mass of silver reaching cells, including ions and NPs, divided by the total volume of the media). At a given experimental time, this target cell dose is lower than the nominal media concentration because of NP transport (e.g. particle settling in static experiments, Brownian motion) and dissolution in the culture media before reaching cells. Only under ideal conditions (t → ∞) are the two concentrations equal. Therefore, in order to carry out more accurate dose-response analyses, minimizing animal testing and avoiding possible misleading conclusions, new computational [10,27,31] and *in vitro* experimental methodologies [32,44,48] have been proposed. These approaches were subsequently combined for better describing the nanotoxicity induced by the direct interactions between Ag NPs and the biological components involved [3]. In this paper we propose a novel integrated *in vitro* and *in silico* pipeline which combines experimental validation of Ag NP sedimentation and dissolution models with the computation of target cell dose (i.e. Ag NPs and dissolved $Ag^+$ in cells) affecting the measured viability of a cell monolayer. Ag NP sedimentation measured in static-cuvette experiments was compared with theoretical values computed by the particokinetic model (ISD3) proposed by Thomas and colleagues [41]. Having verified that the model correctly predicts the sedimentation process, we used it for estimating the total silver concentration perceived by human umbilical vein endothelial cells (HUVECs) and human hepatoma-derived immortalised hepatocyte C3A cells cultured in 96 well plates. Finally, the cell viability was related to the computed target cell doses and compared to the nominal media concentrations of Ag NPs initially administered to cells.

## Materials and Methods

*In vitro – in silico pipeline*

Figure 1 schematises the integrated *in vitro – in silico* pipeline used in this study. We first adapted the ISD3 model (Thomas et al 2018) to our experimental configuration (Fig. 1A).

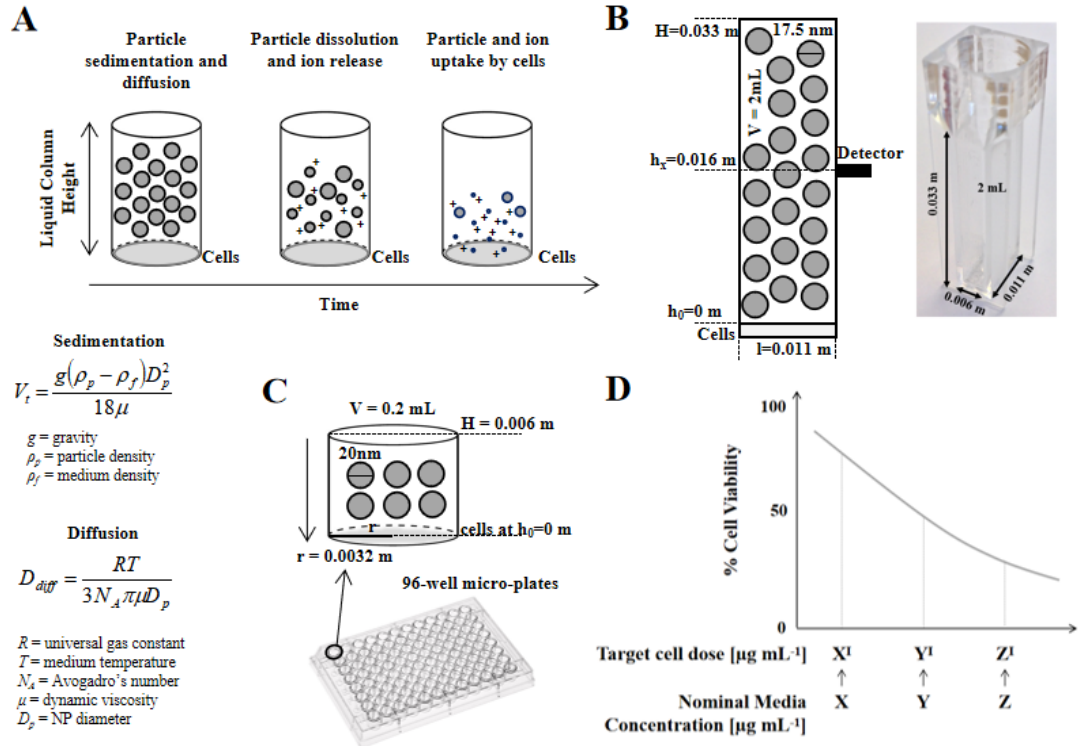

**Figure 1. Particokinetics and experimental validation. (A)** Graphical representation of Ag NP sedimentation, diffusion and dissolution in time as described by ISD3 (Thomas *et al.*, 2018). **(B)** Nanoparticle sedimentation in static-cuvette experiments for model validation. **(C)** 96-well micro-plates experiments for evaluating cell toxicity induced by computed effective Ag NP doses. **(D)** Cell viability vs. computed target cell doses at time *t*. Notably, at a finite experimental time *t*, the computed target cell doses (here denoted with the apostrophe, i.e. X', Y' and Z') are lower than their respective nominal media concentrations (i.e. X, Y and Z).

Then we measured Ag NP sedimentation in static-cuvette experiments and compared the results with those computed by the model (Fig. 1B). In the next step, the model was used to estimate the target cell dose in 96-wells plated with cell monolayers (Fig. 1C). HUVEC and C3A cells were then cultured in 96-microwells and exposed to Ag NPs for up to 24 hours and the cell viability was related to the computed target cell doses as well as to the nominal media concentrations of Ag NPs administered to cells (Fig. 1D).

*Cell cultures*

Human umbilical vein endothelial cells (HUVECs) and human hepatoma-derived immortalised hepatocyte C3A cell line (ATCC Culture, USA) have been obtained by means approved by the appropriate ethical committees and used for evaluating nanotoxicity [44]. Specifically, HUVECs were obtained as described in our previous work [44] and seeded at a concentration of 20,000 cells cm$^{-2}$ on 1% w/v gelatin coated 96-well plates. Cells were allowed to reach confluence (typically 24 h) before exposure to Ag NPs. C3A hepatocytes were seeded at a density of 200,000 cells cm$^{-2}$ on collagen coated plates and incubated for 24 h before experiments. In view of conducting future co-culture experiments in a flow through system, the same medium was used for Ag NP exposure experiments for both HUVEC and hepatocytes. The cells were cultured in Eagle's Minimum Essential Medium (EMEM, Lonza Bioscience, Basel, Switzerland), supplemented with 10% Fetal Bovine Serum (FBS, PAA, Pasching, Austria), 1% Penicillin/Streptomycin/Amphotericin B, 2 mM L-Glutamine, 1% non-essential amino acids, 1% MEM vitamins solution (all from Lonza Bioscience, Basel, Switzerland), 10 µg mL$^{-1}$ Endothelial Cell Growth Supplement (ECGS), 10 ng mL$^{-1}$ Human Epidermal Growth Factor (hEGF), 3 ng mL$^{-1}$ basic Fibroblast Growth Factor (bFGF), 1 µg mL$^{-1}$ Hydrocortisone, and 10 µg mL$^{-1}$ Heparin Sodium Salt (all from Sigma-Aldrich, St. Louis, USA).

*Silver nanoparticles*

Ag NM300 from Ras GmbH, an OECD referenced nanomaterial (NM) with a nominal diameter of 20 nm sonicated in DMEM, was purchased as 10% w/w suspension in a aqueous solution containing 7% v/v ammonium nitrate as stabilizing agent and 4% v/v Tween 20 and 4% v/v polyoxyethylene glycerol trioleate, as emulsifiers. The protocol developed by Klein *et al.* (2011) at the European Commission's Joint Research Centre (JRC) and employed in several publications was used for a stock solution preparation [19,20]. Ag NPs suspended in the medium were experimentally characterized as discussed by JRC (Klein *et al.*, 2011), using the methods reported in Kermanizadeh [22] and Ucciferri (Ucciferri *et al.*, 2014). The methods are summarised in the Supplementary Materials.

*Static sedimentation experiments*

A 2 mL quartz cuvette (Fig. 1B) was filled with the Ag NP suspension and corked to avoid evaporation. Sample absorbance was measured at 414 nm for up to 24 h using a Varian Cary UV spectrophotometer equipped with 1 mm hole positioned at half height of the cuvette. Detected nanoparticles were monitored at different nominal media concentrations (i.e. 5, 10, 15, 40, 50, 80 µg mL$^{-1}$). For each Ag NP concentration used, the absorbance at the beginning of the experiment (time 0 h) was used to generate a calibration curve.

*NP exposure and toxicity assay*

To assess Ag NP toxicity, the medium of HUVEC and C3A cells cultured in 96-well plates (Fig. 1C) was replaced with 0.2 mL of medium containing homogeneously suspended Ag NPs at different concentrations (from 0 to 100 µg mL$^{-1}$) at experimental time 0. Cells were then cultured for up to 24 h. To assess cell toxicity, medium containing Ag NPs or dispersant was removed after each incubation time, and fresh medium and Alamar reagent (CellTiter-Blue® Promega, Madison, USA) were added. Cell viability was obtained from the slope in fluorescence emission within a 2 h time frame measured with a plate-reader spectrophotometer (Omega-Fluostar Inc) and was expressed as a percentage with respect to cells exposed to 0 µg mL$^{-1}$ Ag NP. For each particle concentration the interaction with the assay was finally tested in the absence of cells in order to check for any interference curve.

*In silico model*

We used the particokinetic model for predicting the silver concentration (in the form of Ag NPs and Ag$^+$ ions) perceived by cells in a well. Briefly, ISD3 simulates diffusion and sedimentation of spherical nanoparticles in liquid media as well as the NP dissolution process.

It also models the cellular uptake of both Ag NPs and Ag⁺ ions in the system [41] (Fig. 1A). ISD3 assumes that NPs are instantaneously taken up by cells when reaching their surface, while cell uptake of ions is explicitly modeled as a membrane diffusion process. Kinetics of uniformly distributed ions in liquid media and variations in ion concentrations due to surface adsorption and particle dissolution are finally simulated.

The numerical approach underlying ISD3 is based on Stoke's equation for modelling the sedimentation process and Fick's law for describing the diffusion rate and the sedimentation velocity across the liquid media (Fig. 1A). The dissolution kinetics, estimated by fitting experimental data as a function of time in the medium of interest, are described in Figure 2 and described in [41]. Following the notation and definitions in Thomas et al [41], $k_f$ is the rate constant for the transfer of ions from the particle surface to the free ion state, $k_p$ the rate constant for the slow transfer of ions from the particle surface to the proteins, $k_{p2}$ the rate constant for the initial fast transfer of silver ions from the particle surface to the protein-bound state, $k_{f2p}$ the rate constant for the transfer of free ions from solution to the protein-bound state, and $k_{p2f}$ the rate constant for the transfer of ions from the protein-bound state to the free ion state.

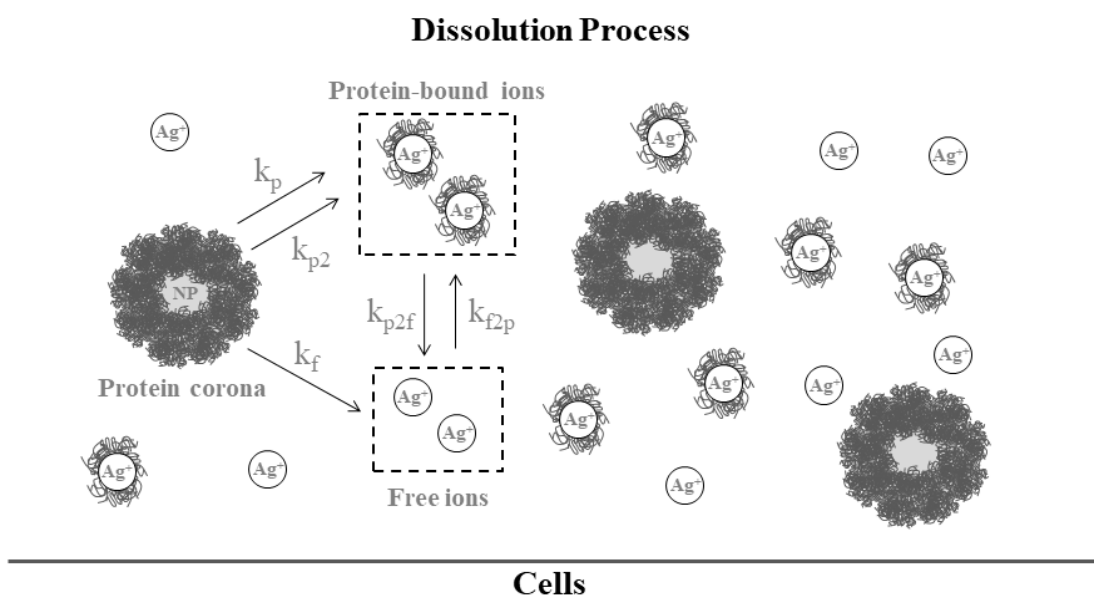

**Figure 2. Schematic of Ag NP dissolution process.** The rate dissolution constants from Ag NPs to free ions ($k_f$), protein ($k_p$) and protein-bound NPs ($k_{p2}$) are described in ISD3. The rate constants for the free ions in solution to protein-bound NPs ($k_{f2p}$) and the transfer of ions from protein-bound NPs to free ions ($k_{p2f}$) are also considered.

Parameters required for running dose calculations, as well as the experimental approaches used for evaluating the input values and additional methods, are listed in Table 1 and 2.

| | PARAMETER | UNIT | INPUT IN LINE | METHODS FOR MEASURING INPUTS (Thomas et al., 2018) | ADDITIONAL METHODS | ADDITIONAL REFERENCES |
|---|---|---|---|---|---|---|
| **Liquid Media Characteristics (lines 50:60)** | Dish depth | [m] | 51 | Directly extracted from the experimental conditions | | |
| | Volume | [mL] | 52 | | | |
| | Temperature | [K] | 53 | | | |
| | Viscosity | [N s m$^{-2}$] | 54 | | | |
| | Density | [g mL$^{-1}$] | 55 | | | |
| **Primary particle characteristics (lines 63:68)** | Hydrodynamic diameter | [nm] | 64 | Measured using Dynamic Light Scattering (DLS) in the medium of interest (e.g., 10% fetal bovine serum (FBS) + RPMI 1640 medium). It is calculated from the intensity weighted average translational diffusion coefficient by using cumulant analysis on the autocorrelation function using vendor provided software. | Fluctuation correlation spectroscopy (FCS) | Dominguenz-Medina et al. 2016 |
| **Protein-coated particle characteristics (lines 70:73)** | Effective density | [g mL$^{-1}$] | 71 | Measured via the volumetric centrifugation method (VCM) in 10% FBS + RPMI. | A sample of nanoparticle suspension can be centrifuged in a packet cell volume (PCV) tube to produce a pellet, the volume of which can be measured and used to estimate the effective density of the nanoparticles in suspension | DeLoid et al. 2015 |
| | Protein corona | [nm] | 72 | The thickness of the protein layers (or protein corona) is assumed to be to ½ the difference in diameters measured by DLS and Transmission Electron Microscopy (TEM), producing an effective particle size equal to the value measured by DLS. | Differential centrifugal sedimentation (DCS); Size exclusion chromatography (SEC); Isothermal calorimetry (ITC); Mass spectroscopy; Circular dichroism (CD); Fluorescence quenching; Surface plasmon resonance (SPR). | Cedervall et al. 2007 Lundqvist et al. 2008 Lacerda et al. 2010 Deng et al. 2010 Walczyk et al. 2010 Duran et al. 2015 |
| **Initial ptcl. size distribution (lines 75:98)** | Effective diameter | [nm] | 78 | The effective diameter (i.e., the average diameter of normal ptcl. size distribution) is measured by combining the hydrodynamic diameter and the protein corona as follows: *Effectve diameter = Hydrodynamic diameter + (2 \* protein corona)* | | |
| **Grid specification Boundary conditions (lines 119:130)** | del_dp | [nm] | 120 | Grid spacing along diameter axis (del_dp) and the number of grid points in the x-axis direction (Nx) take into account the effects of dissolution but the dissolution kinetics are not modelled as a particle surface area limited mechanism. The lowest particle diameter (dp_l) is consistent with the core diameter evaluated via TEM. | | DeLoid et al. 2015 |
| | dp_l | [nm] | 121 | | | |
| | Nx | # | 126 | | | |
| **Time integration inputs (lines 132:136)** | tmaxh | [hours] | 120 | Time integration inputs such as the Total simulation time (tmaxh) and the Number of time steps (Nt) are set depending on the specifications requested by the nanotoxicological study and according to the experimental conditions. | | |
| | Nt | # | 121 | | | |

**Table 1. List of ISD3 input parameters.** 1) Liquid media characteristics; 2) protein-coated particle characteristics; 3) initial particle size distribution; 4) grid specification and boundary conditions; 5) time integration inputs.

| | PARAMETER | UNIT | INPUT IN LINE | METHODS FOR MEASURING INPUTS (Thomas et al., 2018) | ADDITIONAL METHODS | ADDITIONAL REFERENCES |
|---|---|---|---|---|---|---|
| **Dissolution model inputs (lines 139:164)** | $k_f$ | const. | 142 | Rate constants for: a) rate of ion transfer from NP surface to free ion state ($k_f$); b) slow ion transfer rate from NP surface to dissolved proteins ($k_p$); c) initial fast Ag ion transfer rate of silver ions from NP surface to the protein-bound state ($k_{p2}$); d) the transfer of free ions from solution to protein-bound NPs ($k_{f2p}$); e) the transfer of ions from the protein-bound NPs to the free ion state ($k_{p2f}$). These rate constants, as well as the saturated concentration of free ions in solution ($C_{sat}$), the concentration of binding sites available on the proteins per FBS for the slow binding period of the ions ($n$) and the concentration of binding sites available on the proteins per FBS for the fast binding period of the ions ($n_2$), were estimated by fitting experimental data measured in cell culture media with the differential equations describing the free ion concentration (Eq. 9 in Thomas et al.) and the protein-bound NP ion concentration (Eq. 10 in Thomas et al.). In this regard, the silver levels were experimentally quantified by using inductively coupled plasma-mass spectrometry (ICP-MS). | Ion release can be also measured by Atomic Absorption Spectroscopy (AAS) | DeLoid et al., 2015 Liu et al., 2015 Böhmert et al., 2018 |
| | $k_p$ | const. | 147 | | | |
| | $k_{p2}$ | const. | 150 | | | |
| | $k_{f2p}$ | const. | 153 | | | |
| | $k_{p2f}$ | const. | 155 | | | |
| | $C_{sat}$ | [µg mL$^{-1}$] | 145 | | | |
| | $n$ | [µg (mL%FBS)$^{-1}$] | 158 | | | |
| | $n_2$ | [µg (mL%FBS)$^{-1}$] | 161 | | | |
| **Cellular uptake of silver ions (lines 166:172)** | dis2 | [cm] | 169 | Inputs for cellular uptake of silver ions: a) the thickness of cell membrane (dis2); b) the cell volume (cellv); c) the media silver ion partition coefficient (pc2l); d) the diffusion coefficient (dl2). The media silver ion partition coefficient and the diffusion coefficient were estimated by fitting the silver levels measured in cells by using ICP-MS with the final differential equations describing the free ion concentration (Eq. 12 in Thomas et al.) and the protein-bound NP ion concentration (Eq. 13 in Thomas et al.). The thickness of cell membrane and the cell volume used by Thomas and co-workers were specific for RAW 264.7 macrophage cells. | | |
| | dl2 | cm$^2$ h$^{-1}$ | 170 | | | |
| | pc2l | | 171 | | | |
| | cellv | [mL] | 172 | | | |
| **Additional information:** options for the particle agglomerate characteristics are available at lines 99:104 | | | | | | |

**Table 2. Additional ISD3 input parameters.** Dissolution model inputs and Cellular uptake of silver ions.

The input values used in [41] correspond to those for a nominal 20 nm silver nanomaterial in 10% serum. Hence, they are also suitable for Ag NMP300 in 10% FBS in the concentration range used in our studies (5-80 µg mL$^{-1}$) and only a few values (shown in bold in Table 3) were modified for modelling the dissolution process involved in our static-cuvette and 96-well micro-plates. Should other particles be used, they must first be characterized using the methods suggested in Tables 1 and 2.

|  | Parameters | Unit | Cuvette | 96-well micro-plate |
|---|---|---|---|---|
| **Media Features** | **Height** | [m] | 0.033 | 0.006 |
|  | **Volume** | [mL] | 2 | 0.2 |
|  | Temperature | [K] | 310 | 310 |
|  | Viscosity | [N s m$^{-2}$] | 0.00074 | 0.00074 |
|  | Density | [g mL$^{-1}$] | 1 | 1 |
|  | **Surface area** | [m$^2$] | 6.6 10$^{-5}$ | 3.3 10$^{-5}$ |
| **Particle Features** | **Particle Size Diameter** | [nm] | 17.5 | 17.5 |
|  | Primary Particle Density | [g cm$^{-3}$] | 10 | 10 |
|  | **Protein Layer Thickness** | [nm] | 51.25 | 51.25 |
|  | **Effective Particle Diameter** | [nm] | 120 | 120 |
|  | Effective Density | [g cm$^{-3}$] | 1.454 | 1.454 |
| **Grid Spacing, Time and Particle Dissolution** | **Grid Spacing along Particle Diameter** | [nm] | 0.5 | 0.5 |
|  | **Number of High Grid Spacing** | - | 100 | 100 |
|  | **Time** | [h] | 24 | 24 |
|  | Dissolution of Particles to free ions | [mL nm$^{-2}$ h$^{-1}$] | 0.0006 10$^{-14}$ | 0.0006 10$^{-14}$ |

**Table 3. Particokinetic model inputs.** Parameters adapted for modelling dissolution processes in static-cuvette and 96-well micro-plate experiments in this study are shown in bold.

*Data processing*

In order to evaluate sedimentation rates static-cuvette experiments were performed in duplicate, while 96-well plate *in vitro* experiments were performed in triplicate. Data were processed and the dose-response curves plotted using MATLAB R2013b (Mathworks). Results are expressed as means ± standard deviation.

**Results**

The Ag NPs used in this work have been extensively characterized (Klein *et al.*, 2011; Kermanizadeh *et al.*, 2013; Ucciferri *et al.*, 2014) in the medium used in our experiments. The particles have an average size of 17.5 nm with euhedral morphology as measured using transmission electron microscopy (TEM). Finally, the mean effective diameter observed using Nanosight system is 120 nm with monomodal size distribution. Data are summarized in Table 4.

| Ag NPs | Nominal Diameter (nm) | TEM | | Nanosight analysis | |
|---|---|---|---|---|---|
|  |  | Average size (nm) | Morphology | Effective diameter in medium (nm) | Size distribution (PTA analysis) |
|  | 20 | 17.5 | euhedral | 120 ± 4 | monomodal |

**Table 4. Ag NP characterization.** Characterization performed by TEM and single particle tracking analysis (PTA) in the cell culture medium.

Experimentally, we first tracked the Ag NP dissolution and sedimentation process in static-cuvette experiments as a function of the initial administered concentrations (nominal media concentrations). The rationale was to validate the particokinetic model ISD3 provided by Thomas et al. (2018) quantifying how the nano-particle concentrations computationally predicted by the model fit with those measured at half height of the cuvette (Fig. 3; grey bars). A high Pearson coefficient value (r=0.9991) was observed, indicating a strong correlation between predicted and detected concentrations (inset), confirming the reliability of the ISD3 in modelling the particle sedimentation process.

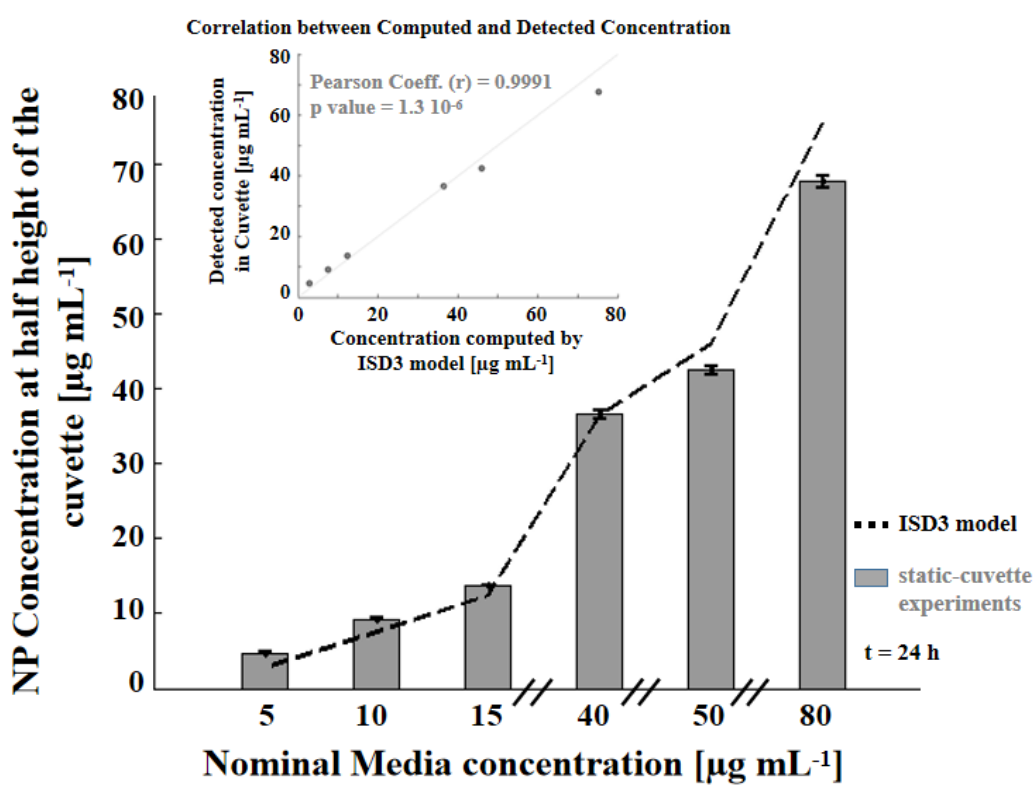

**Figure 3. Validation of the particle sedimentation process in static-cuvette experiments.** Ag NP concentrations at half height of the cuvette (grey bars) fit well with those predicted by the particokinetic model (black dashed line). The scatter plot (inset) shows a strong correlation (r = 0.9991) between the experimental and computed Ag NP concentrations.

Subsequently, we compared the target cell doses predicted by the ISD3 model with increasing nominal media concentrations administered to HUVEC and C3A cells cultured in 96-well micro-plates. The rationale was to test the assumption that the total silver concentrations coming into contact with the cellular system (i.e. the target doses) are lower than the nominal media concentrations because of the particle dissolution and sedimentation processes in media. In particular, we computed the target cell doses (Fig. 4A, solid lines) for different Ag NP nominal media concentrations (initially administrated to cells, dashed lines) over 24 hours. At the end of our experiment (24 h), the target cell doses were much lower than their respective nominal media concentrations as represented by the vertical distance between the black and grey lines in Figure 4B.

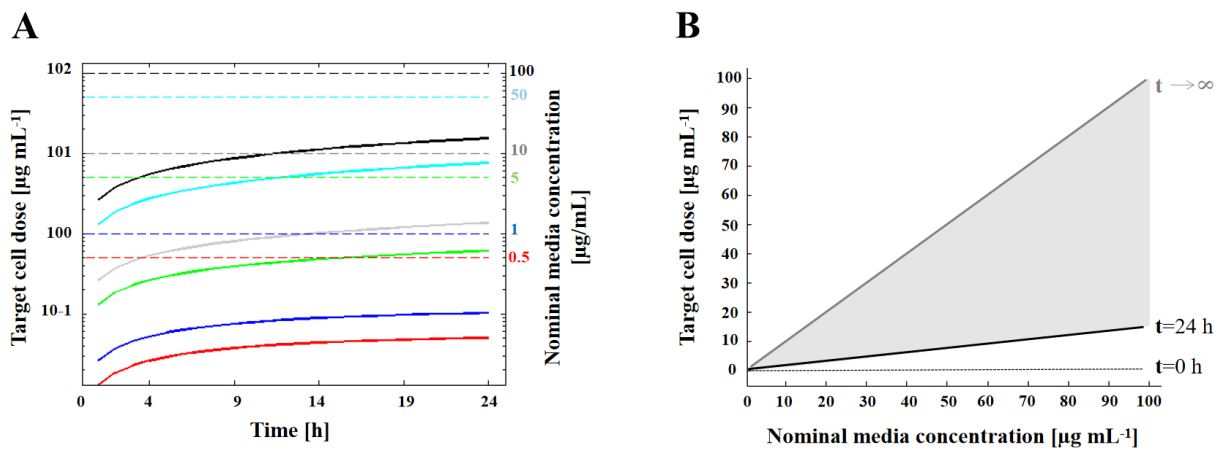

**Figure 4. Target cell doses computed over 24 hours. (A)** Target cell doses (solid lines) as a function of time and computed (dashed lines) from 6 nominal media concentrations (red: 0.5 µg mL$^{-1}$; blue 1 µg mL$^{-1}$; green 5 µg mL$^{-1}$; grey 10 µg mL$^{-1}$; cyan 50 µg mL$^{-1}$; black 100 µg mL$^{-1}$. Target cell doses are equal to zero at t = 0 (not apparent in the semi-log pot). **(B)** Target cell doses computed at t = 24 h (black solid line) versus nominal media concentrations. Notably, only under ideal conditions (t → ∞) are these two concentrations equal (grey solid line). The black dotted line denotes target cell doses at t = 0 h, which are all null regardless of the nominal media concentration.

The differences between them, here expressed as Δ = (nominal media concentration - target cell dose)/(nominal media concentration), are reported in Table 5. As expected, the target cell doses were proportional to their respective nominal media concentration (Table 5).

| Nominal media conc. [µg mL$^{-1}$] | Ion in cells [µg mL$^{-1}$] | Particle in cells [µg mL$^{-1}$] | Target cell dose [µg mL$^{-1}$] | Δ [%] |
|---|---|---|---|---|
| 0.5 | 0.003 | 0.047 | 0.05 | 89.99 |
| 1 | 0.006 | 0.096 | 0.103 | 89.73 |
| 5 | 0.018 | 0.594 | 0.612 | 87.75 |
| 10 | 0.02 | 1.341 | 1.361 | 86.38 |
| 50 | 0.032 | 7.612 | 7.644 | 84.71 |
| 100 | 0.042 | 15.495 | 15.537 | 84.46 |

**Table 5. Nominal media concentrations vs. Target cell doses.** Target cell doses (i.e. Ag NPs and dissolved Ag$^+$ in cells) are about 85% lower than their respective nominal media concentrations (i.e. the initial administered doses) at 24 h.

These results demonstrate how the nanotoxicity potentially induced by the NP exposure could be misunderstood if related to the initial administered concentrations. Specifically, relating cell viability at a given experimental time to the nominal media concentration leads to an underestimation of the nanoparticle toxicity. Finally, no toxic effects were observed for Ag NP dispersant, nor did Ag NPs interfere with the viability assay (data not shown).

Having evaluated the target cell doses during the entire exposure period (i.e., from 0 to 24 h) and quantified differences with the nominal media concentrations (Δ values), we measured HUVEC and C3A viability at 6 h (Fig. 5A), 16 h (Fig. 5B) and 24 h (Fig. 5C) of culture as a function of the computed target cell doses.

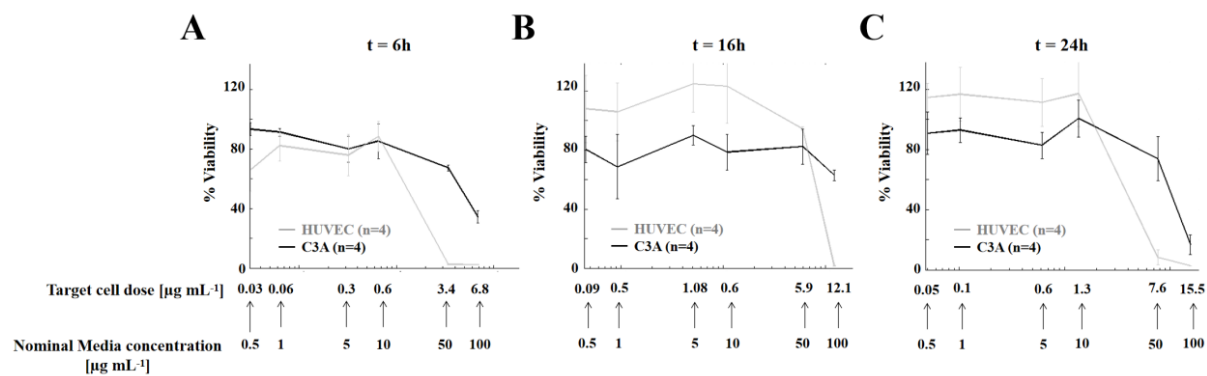

**Figure 5. HUVEC (gray line) and C3A (black line) viability as a function of increasing target cell doses.** Panels show the viability as measured by the Alamar blue assay at 6 h **(A)**, 16 h **(B)** and 24 h **(C)**. Data shown as mean ± std (n=4).

The rationale was to better interpret time-dependent nanotoxicity effects induced by the effective silver concentration coming into contact with cells. Both HUVEC (gray line) and C3A (black line) viability decreased with increasing target (and nominal) cell dose. From the graphs it is also evident that the Δ has implications on the estimation of IC50 values. In fact, referring to the nominal media concentration, the IC50 value at 24 h is estimated as 41.04 µg mL$^{-1}$ for HUVEC and 54.78 µg mL$^{-1}$ for C3A. However, if one considers the target cell dose (actually experienced by cells), the IC50 drops significantly (~80%): 8.13 µg mL$^{-1}$ for HUVEC and 10.95 µg mL$^{-1}$ for C3A.

**Discussion**

In this work we propose a novel integrated approach based on computational calculations and *in vitro* experiments to provide more accurate dose-response analyses and to minimize time-consuming, expensive and ethically sensitive *in vivo* tests [50]. Having verified the accuracy of the ISD3 model through sedimentation experiments, we evaluated the viability of HUVEC and C3A cells cultured in 96-well plates as a function of the total silver mass reaching cells over time (namely, target cell dose). In this regard, much uncertainty still remains regarding which constituent (i.e., NPs, free and protein-bound ions) contributes to cellular toxicity [2,36]. Bouwmeester and co-workers [4], as well as Wang and colleagues [47], suggest that exposure to silver ions formed extracellularly is responsible for observed toxic effects. Conversely, other groups assume that the internalized nanoparticles (i.e. NPs taken up by cells) undergo rapid dissolution resulting in silver ions inducing toxicity [17,40]. The target cell dose includes both dissolved ions and particles coming into contact with cells (i.e. Ag NPs + dissolved Ag$^+$ in cells) and therefore can account for both NP and ion related toxic effects.

Our results show that the target cell doses computed at 24 h were around 85% lower than their corresponding nominal media concentrations initially administered to cells (t=0 h). Moreover, both HUVEC and C3A viability decreased with concentration and with the time of exposure. Thus, Ag NP toxicity reported in the form of nominal dose-response data is likely to be significantly underestimated, and the extent of error increases significantly at short exposure times.

The present study highlights some fundamental concerns in *in vitro* nanotoxicology and suggests that a number of issues still need to be addressed before data from cell culture experiments can be considered reliable for predicting NP toxicity.

As far as the model is concerned, for one it assumes the Ag NPs to be immediately absorbed once the cell surface is reached. Furthermore, ISD3 does not simulate the dissolution process of the Ag NPs once they have been taken up by the cells and it is limited to a 2 dimensional cell configuration, which is not considered "physiologically relevant". By modelling the kinetics involved within the three-dimensional geometries of the cellular constructs, we should be able to predict the Ag NP uptake at multiple layers in a more realistic *in vivo*-like configuration. Moreover, the computed target cell doses sequentially absorbed "layer by layer" would allow an even more accurate analysis of the effects of nanotoxicity. In this regard, the three-dimensional and scaffold-less cellular aggregates in the form of "organoids", "microtissues" or "spheroids" which resemble the cytoarchitectural arrangement of the human organs (e.g., liver, lung and specific brain regions) are currently being used as more physiologically relevant *in vitro* systems for experimental validation [28]. For example, Kermanizadeh and co-workers have investigated three-dimensional human liver microtissue models, exposing them to different types of nanomaterials (including Ag NPs). They used Ag NP concentrations within the range employed here, demonstrating that their repeated exposure is more damaging to the liver tissue in comparison to a single exposure [21].

In addition, for the model configuration described in our work, we would need to compute and experimentally validate not only the NPs diffusing towards the cells cultured on the bottom of the experimental set-up, but also those particles coming into contact with the cells possibly adjacent the walls. To the best of our knowledge, this second computational aspect has been exclusively provided by Bohemert and co-workers (2018). They indeed modelled the dosimetry and the exposure to nanoparticles of Caco-2 cells forming a confluent monolayer on the bottom of the cell culture dish, but also growing up the encircling wall of the cell culture dish [3].

Another aspect to consider is that Ag NP sedimentation and cell toxicity were modelled and experimentally evaluated under static conditions. It would be interesting to extend the combined computational-experimental approach to dynamic experiments, implemented with a modular bioreactor similar to that described in our previous work [46] or a microfluidic system [51]. A multiphysics computational approach could be used to model the combined effects of particle sedimentation, diffusion, dissolution and convection as a function of the involved fluxes characterizing the dynamics of this type of more "physiologically relevant" *in vitro* model.

Finally, it would be interesting to quantify the cell viability affected by absorbed nanoparticles of different materials and diameter sizes. These features, included the protein corona on the surface of the nanoparticles (in this work protein layer equal to (120-17.5) / 2 = 51.25 nm), model the primary particle structure and specifically influence the induced nanotoxicity effects [3,41].

In conclusion, the combination of computational and experimental approaches could provide more reliable dose-response analyses and hazard assessment. This study paves the way for the development of a methodological framework which combines the expertise of modellers, toxicologists and scientists studying in nanotechnology and nanoscience to enable good experimental design and more accurate in vitro to in vivo data extrapolation.

**Acknowledgments**

The work leading to this paper has received funding from the European Union's H2020 research and innovation programme under grant agreement No. 760813 (PATROLS).

**Disclosure**

The authors have no relevant interests to disclose.

**Data availability statement**

Data are available from the corresponding author upon reasonable request.

**Author contributions**

DP analyzed the data. NU performed experiments. DP, GM and AA wrote the paper and interpreted the data. DP and AA edited and prepared the final layout. All authors gave final approval of the paper.

# Supplementary Materials

# An integrated *in vitro* – *in silico* approach for silver nanoparticle dosimetry in cell cultures

Daniele Poli[1], Giorgio Mattei[2], Nadia Ucciferri[1] and Arti Ahluwalia[1, 2]

[1]Research Center E. Piaggio, University of Pisa, Pisa, Italy; [2]Department of Information Engineering, University of Pisa, Pisa, Italy

### Ag NP characterization

Particles were dispersed in the complete cell culture medium. The same dispersion protocol was followed for all exposure experiments and when characterizing possible agglomeration tendencies and stability of dispersions. Ag NPs were stored as viscous liquid at the concentration of 10% (w/w) dispersed in water (75%) with a stabilizing agent (ammonium nitrate 7%) and emulsifiers (Tween 20 and polyoxyethylene glycerol trioleate, 4% each). For preparation of a stock solution we used the protocol provided by the ENPRA project (www.enpra.eu): ~100 mg of silver particle solution was weighed and dispersed in ~38 mL of water supplemented with 2% FBS to a final concentration of 2.56 mg mL$^{-1}$. Immediately after preparation the stock solution was sonicated for 16 min in an ultrasonic bath (Bandelin electronic, Berlin, Germany) at 200 W. Dilutions between 0 and 80 µg mL$^{-1}$ were prepared in the complete cell culture medium. Before use samples were again sonicated for 15 min.

Particle hydrodynamic diameter was determined via dynamic light scattering using Malvern Zeta sizer nano (Malvern Instruments, Herrenberg, Germany). Size and polydispersity were determined at time points 0, 1, 2, 4, 6, 8 and 24 hours after preparation of the respective dilution. Between measurements, dilutions were kept at room temperature and protected from light. Diluted samples were again sonicated for 16 minutes in an ultrasonic bath. For each time point, two independent samples were measured three times each and mean and standard deviation are reported.

Size distribution and stability were also investigated using NanoSight LM10 instrument (NanoSight Limited, Amesbury, UK). The system is based on dynamic light scattering but uses single particle tracking analysis. At each time point, three measurements were performed; mean value and standard deviation were calculated. Here, sample data at 0, 8 and 24 h are reported (Figure S1).

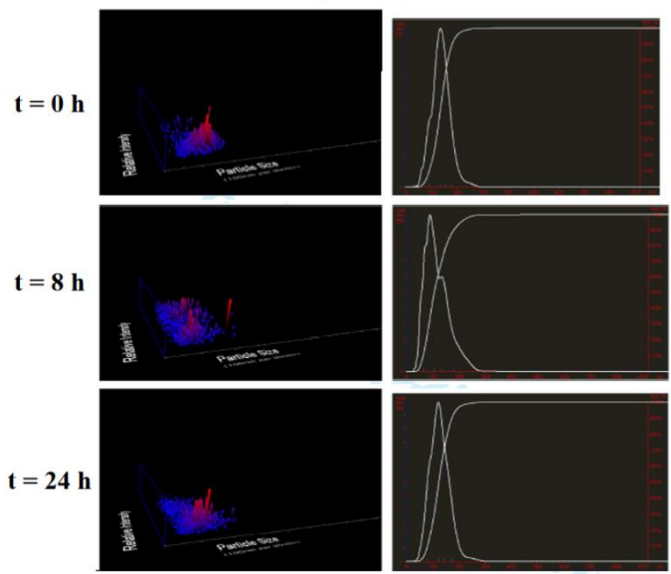

**Figure S1.** Ag NPS tracking at 0 h, 8 h and 24 h.

**Running the ISD3 model**

The MATLAB code of ISD3 model is available from Thomas et al. (2018) and can be downloaded at https://nanodose.pnnl.gov. The boundary conditions of the experimental system can be easily updated by modifying the input data defined and discussed within the MATLAB script named "inputdata.m". Right clicking and run the "isd3.m" file for beginning the simulations. Outputs are named as shown in Table S1 and saved into a MATLAB or EXCEL file. The name of this file can be entered into "inputdata.m" at line 48.

| | OUTPUT (into EXCEL file) | VARIABLE (in "isd3.m" file) | UNIT | DEFINITION | VARIABLE IN LINE |
|---|---|---|---|---|---|
| **NANOPARTICLES** | LiquidMediaParticleNumber | output.t_partNumL | | Number of particles in liquid media vs. time | 1078 |
| | CellDepositedParticleNumber | output.t_PartNumC | | Number of particles deposited on cells vs. time | 1079 |
| | TotalParticleNumber | output.t_partNumTot | | Total number of particles in the system vs. time | 1080 |
| | LiquidMediaParticleConc | output.t_partConcL | [µg mL$^{-1}$] | Mass concentration of particles in liquid media vs. time | 1207 |
| | | output.tx_partConcL | [µg mL$^{-1}$] | Mass concentration of particles in liquid media vs. time and x | 1191 |
| | CellDepositedParticleConc | output.t_partConcC | [µg mL$^{-1}$] | Mass concentration of particles deposited on cells vs. time | 1278 |
| | LiquidMediaParticleMass | output.t_partMassL | [µg] | Mass of particles in liquid media vs. time | 1201 |
| | CellDepositedParticleMass | output.t_partMassC | [µg] | Mass of particles deposited on cells vs. time | 1277 |
| | LiquidMediaParticleSurfArea | output.t_partSAL | [cm$^2$] | Surface area of particles in liquid media vs. time | 1235 |
| | CellDepositedParticleSurfArea | output.t_partSAC | [cm$^2$] | Surface area of particles deposited on cells vs. time | 1279 |
| **IONS** | LiquidMediaFreeIonConc | output.t_ionConcFreeL | [µg mL$^{-1}$] | Mass concentration of ions in free state in liquid media vs. time | 1286 |
| | LiquidMediaProteinBoundIonConc | output.t_ionConcBndL | [µg mL$^{-1}$] | Mass concentration of ions in protein-bound state in liquid media vs. time | 1287 |
| | LiquidTotalIonConc | output.t_ionConcTotL | [µg mL$^{-1}$] | Total concentration of ions in liquid media vs. time | 1288 |
| | CellIonConc | output.t_ionConcC | [µg mL$^{-1}$] | Mass concentration of ions deposited in cells vs. time | 1289 |
| | LiquidMediaFreeIonMass | output.t_ionMassFreeL | [µg] | Mass of ions in free state in liquid media vs. time | 1292 |
| | LiquidMediaProteinBoundIonMass | output.t_ionMassBndL | [µg] | Mass of ions in protein-bound state in liquid media vs. time | 1293 |
| | LiquidTotalIonMass | output.t_ionMassTotL | [µg] | Total mass of ions in liquid media vs. time | 1294 |
| | CellIonMass | output.t_ionMassC | [µg] | Total mass of ions in cells vs. time | 1295 |
| | totalConc | output.t_totalConc | [µg mL$^{-1}$] | Total concentration (particles + ions) in liquid media and in cells vs. time | 1298 |

**Table S1.** List of the outputs generated by ISD3 particokinetic model and saved in an EXCEL file. The mass concentration of nanoparticles in liquid media vs. time and height x (i.e. output.tx_partConcL) is used for extracting the predicted Ag. NP concentration computed at half height of the cuvette (dashed line, Fig. 3). The target cell doses computed over 24 h (Fig. 4 and 5) are obtained from the variables named output.t_partConcC and output.t_ionConcC which are saved into the EXCEL file as "CellDepositedParticleConc" and "CellIonConc", respectively.